\documentclass[twocolumn,prd,aps,showpacs,showkeys,amsmath,amssymb]{revtex4-1}
\usepackage{bm}
\usepackage{color}
\usepackage{amsmath}
\usepackage{amsfonts}
\usepackage{verbatim}
\usepackage{amssymb}
\usepackage{graphicx}
\usepackage{epstopdf}
\usepackage{mathrsfs}
\usepackage{epsfig}
\usepackage{slashed}
\usepackage{bbold}

\providecommand{\U}[1]{\protect\rule{.1in}{.1in}}
\newcommand{\bl}{\boldsymbol}

\begin{document}

\title{Quasinormal Modes in Generalized Nariai Spacetimes}
\author{Jo\'{a}s Ven\^{a}ncio and Carlos Batista}
\email[]{joasvenancio@df.ufpe.br, carlosbatistas@df.ufpe.br}
\affiliation{Departamento de F\'{\i}sica, Universidade Federal de Pernambuco,
Recife, Pernambuco  50740-560, Brazil}

\begin{abstract}
The perturbations of fields with spin 0, 1/2, and 1 propagating in a higher-dimensional generalization of the charged Nariai spacetime are investigated. The boundary conditions leading to quasinormal modes are analyzed and the quasinormal frequencies are analytically obtained.
\end{abstract}
\maketitle

\section{Introduction}

It is well-known that a string of a guitar produces a characteristic sound when someone hits it. This characteristic sound is the natural way the system finds to respond to the external excitation. Interestingly, similar phenomena are ubiquitous in dynamical systems that are in an equilibrium state. These systems typically respond to a perturbation by oscillating around the equilibrium configuration with a set of natural frequencies, known as the normal frequencies. In particular, when some specific frequency is selected we say that the system is in a normal mode.

Now, do black holes have a characteristic ``sound'' as well? The answer is yes. Studying scattering in Schwarzschild geometry, Vishveshwara found that the evolution of perturbations is given by damped oscillations with natural frequencies that do not depend on the details of the excitation \cite{Vishveshwara}. Since these perturbations decay exponentially in time, they are characterized by complex frequencies. Hence, they are called quasinormal frequencies (QNFs), and the configurations with a single frequency are the quasinormal modes (QNMs) \cite{Press, Ferrari}. The real part of a QNF is associated with the oscillation frequency of the perturbation, while the imaginary part is related to its decay rate. This damping stems from the existence of an event horizon, which prevents incoming signals to be reflected back, yielding dissipation. The interesting fact is that these frequencies depend on the charges of the black hole, such as mass, electric charge, and angular momentum.  Therefore, the measurement of QNFs can be used to obtain the charges of astrophysical black holes \cite{Vishveshwara, Andersson}. This has incited a wide effort to find the QNFs of several gravitational configurations, with several numerical and analytical techniques being devised \cite{Kokkotas, Berti09, Cardoso03, Nollert99}. The interest in QNMs has been renewed by the recent detection of gravitational waves \cite{Abbott}, since now the QNFs are closer of being experimentally accessible. Another reason for studying QNMs is that we would expect, in light of Bohr's correspondence principle,  that they should give some hint about quantum gravity \cite{Hod}. Indeed, a connection between QNFs and the quantization of event horizon area has been put forward \cite{Dreyer,Maggiore:2007nq,Domagala:2004jt}.

From the theoretical point of view, most of the recent works featuring QNMs are concerned with higher-dimensional spacetimes \cite{Konoplya:2011qq,FrolovQNM,ZhidenkoQNM}. For instance, QNMs are used to test the stability of the of certain solutions, this is particularly useful in dimensions greater than four, in which case there is no uniqueness theorem for black holes, so that the stability may be the criteria to select physical configurations among several gravitational solutions \cite{Zhidenko, Liu}. There are several motivations for studying gravitational configurations in dimensions greater than four. For example, string theory, which intends to describe the fundamental interactions of nature in a unified scheme, requires the spacetime to have $10$ dimensions \cite{Mukhi:2011zz}. Actually, there are many other theories that seek to explain our Universe through the use of higher-dimensional theories, for reviews see \cite{Emparan, Csaki:2004ay}. Another source of interest in higher-dimensional spacetimes is the AdS/CFT correspondence, which provides tools to tackle field theories living in $d$ dimensions by means of studying gravitational solutions in $d+1$ dimensions \cite{Maldacena:1997re,Horowitz:2006ct,Hubeny:2014bla}. Through AdS/CFT correspondence, QNFs can be associated to the thermalisation of perturbations in finite temperature field theories \cite{Horowitz, Sachs, Nunez, Keranen:2015mqc, David:2015xqa}.

With the above motivations in mind, in the present article we shall consider a higher-dimensional generalization of the charged Nariai spacetime \cite{Carlos-Nariai} and investigate the dynamics of perturbations of test fields with spins 0, 1/2 and 1. In particular, we investigate the boundary conditions that lead to QNMs and analytically obtain the spectrum of QNFs. The background used here is the direct product of two-dimensional spacetimes of constant curvature, $dS_2 \times S^{2} \times \cdots \times S^{2}$, while the most known higher-dimensional generalization of Nariai spacetime is given by $dS_2\times S^{D-2}$ \cite{Kodama:2003kk,Cardoso:2004uz}. One interesting feature of the spacetime considered here is that it supports magnetic charges besides the electric charge  \cite{Carlos-Nariai}, which lead to a reacher physics. Moreover, spaces that are the direct product of two-dimensional spaces can also be of relevance to model internal spaces in string theory compactifications \cite{Brown}.

The outline of the article is the following. The next section provides a general discussion on the perturbations of matter fields in fixed backgrounds and present the spacetime considered here. Then, in Sec. \ref{Sec.Boundary Cond}, we analyse physically motivated boundary conditions for the background studied here and define four boundary conditions that will be used in the sequel. Sec. \ref{Sec.Potential} then paves the way for the following sections by focusing on the integration of a Schr\"{o}dinger-like equation that will appear on the perturbation of all matter fields. The analysis of the asymptotic forms of the solutions is also performed. In Secs. \ref{Sec.Scalar}, \ref{Sec.Maxwell} and \ref{Sec.Spinor} we treat the perturbations on scalar, Maxwell and Dirac fields, respectively. In these sections the degrees of freedom of these fields are separated, so that the dynamics boils down to  a single differential equation. The boundary conditions are imposed and the QNFs obtained.  Finally, Sec. \ref{Sec.Conclusion} sums up the results and give some perspective on future work. We also provide two appendices, one motivating the ansatz for the Maxwell field that allows separation of the degrees of freedom, App. \ref{Appendix.Maxwell}, and the other describing how the angular part of the spinorial field is tackled, App. \ref{Appendix.Angular}.

\section{Presenting the Problem}\label{Sec.Problem}

In $D$ dimensions, the dynamics of general relativity in spacetimes with cosmological constant $\Lambda$ is described by the Einstein-Hilbert action
\begin{equation}
S\,=\,\frac{1}{16\pi} \int d^{D}x\sqrt{\left | g \right |}\left[ \mathcal{R}-(D-2)\Lambda \right] \,+\,S_{m},
\end{equation}
where $S_{m}$ stands for the action of the matter fields $\{\Phi_i\}$ coupled to gravity. The least action principle allows us to find the equations of motion for the fields $g_{\mu\nu}$ and $\Phi_{i}$ which are given, respectively, by
\begin{align}
 &\mathcal{R}_{\mu\nu} + \frac{1}{2}\left[ \Lambda(D-2)  -\mathcal{R} \right]\,g_{\mu\nu} = 8\pi T_{\mu\nu} \,, \label{EE}\\
 & \frac{\delta S_{m}}{\delta\Phi_{i}}\,=\,0 \label{FE}\,,
\end{align}
where the symmetric tensor $T_{\mu\nu}$ is the stress-energy tensor associated to matter fields defined by
$$T^{\mu\nu}\,=\,\frac{2}{\sqrt{\left | g \right |}} \frac{\delta S_{m}}{\delta g_{\mu\nu}}\,. $$

Now, let the pair $g^{0}_{\mu\nu}$ and $\Phi^{0}_{i}$ be a solution for the equations of motion \eqref{EE} and \eqref{FE}. Then, in order to study the perturbations around this solution, we suppose that our fields are given by
\begin{equation}\label{Pertubation}
g_{\mu\nu}\,=\,g^{0}_{\mu\nu}\,+\,h_{\mu\nu} \quad , \quad  \Phi_{i} \,=\,\Phi^{0}_{i}\,+\,\phi_{i} \,,
\end{equation}
where $h_{\mu\nu}$ and $\phi_{i}$ are assumed to be ``small''. Thus, plugging the ansatz \eqref{Pertubation} into \eqref{EE} and \eqref{FE} and neglecting quadratic and higher order powers of the perturbation fields,  we are left with a set of linear equations satisfied by $h_{\mu\nu}$ and $\phi_{i}$. In general, these equations are coupled, namely $\phi_{i}$ is a source for $h_{\mu\nu}$ and vice-versa. However, in the special case in which
$\Phi^{0}_{i} = 0$ it follows that the equations for  $\phi_{i}$ and $h_{\mu\nu}$ decouple, due to the fact the energy momentum tensor $T_{\mu\nu}$ just have quadratic and higher order powers of the matter field. In other words, when $\Phi^{0}_{i}=0$ we have $T_{\mu\nu}=0$ at first order on the perturbation. In such a case, the dynamics of generic small perturbations of the matter fields is equivalent to studying the test fields $\phi_{i}$ in the fixed background $g^{0}_{\mu\nu}$.

From now on, let us consider matter fields propagating in the background described in Ref \cite{Carlos-Nariai}, a higher-dimensional generalization of the Nariai spacetime whose metric in $D=2d$ dimensions is formed from the direct product of the de Sitter space $dS_{2}$ with $(d-1)$ spheres $S^{2}$ possessing different radii $R_{j}$, namely
\begin{equation}\label{nariai-metric}
g^{0}_{\mu\nu}dx^{\mu}dx^{\nu} = -f(r)dt^{2} + \frac{1}{f(r)}dr^{2}+ \sum_{j=2}^{d}R_{j}^{2}\,d\Omega_{j}^{2} \,,
\end{equation}
where $d\Omega_{j}^{2}$ is the line element of the $j$-th unit sphere and $f(r)$ is a function of the coordinate $r$
\begin{equation}\label{functionf1}
 d\Omega_{j}^{2}\,=\,d\theta_{j}^{2}\,+\, \text{sin}^{2}\theta_{j}\,d\phi_{j}^{2} \quad , \quad f(r)\,=\,1-\frac{r^{2}}{R_{1}^{2}}  \,.
\end{equation}
The radius $R_{1}$ and $R_{j}$ are constants given by
\begin{equation}\label{radii}
    \begin{array}{ll}
      R_{1}&\,=\,\left[\Lambda \,-\,\frac{1}{2}Q_{1}^{2}\,+\,\frac{Q}{2(D-2)} \right]^{-1/2} \,, \\
R_{j}&\,=\,\left[\Lambda \,+\,\frac{1}{2}Q_{j}^{2}\,+\,\frac{Q}{2(D-2)} \right]^{-1/2} \,,
    \end{array}
 \end{equation}
with $Q_{1}$  and $Q_{j}$ being the electric and magnetic charges, respectively, while $Q$ is defined by	
\begin{equation}
Q \,\equiv\, Q_{1}^{2} \,-\, \sum_{j=2}^{d}Q_{j}^{2} \,.
\end{equation}
That is a locally static solution of the equation \eqref{EE} in the presence of the electromagnetic gauge field
\begin{equation}\label{gauge-field}
\bl{\mathcal{A}}^0 \,=\,Q_{1}\, r\, dt \,+\, \sum_{j=2}^{d}Q_{j}R^{2}_{j} \,\text{cos}\theta_{j}\,d\phi_{j} \,.
\end{equation}
One may notice that the Killing vector $\partial_{t}$ is light-like at  the closed null surfaces $r = \pm R_{1}$, so that $r = \pm R_{1}$ are Killing horizons.

The so-called quasinormal modes (QNMs) are solutions of the perturbation equations satisfying specific boundary conditions \cite{Berti09, Kokkotas, Natario, Ortega, Zhidenko}. For instance, in asymptotically flat black hole spacetimes, the boundaries are generally chosen to be the outer event horizon and the infinity, with the perturbation field assumed to be ingoing at the horizon and outgoing at infinity. The eigen-frequencies of this problem are complex, with the imaginary part accounting for the dissipative nature of the event horizon.

In what follows, we shall investigate the QNMs of matter fields of spin 0, 1/2 and 1 in the background \eqref{nariai-metric}. We shall use as the boundaries of this space the horizons $r=\pm R_1$ and consider four types of boundary conditions, as described in the following section. Since the spacetime considered here is the direct product of the de Sitter spacetime with several spheres, it is not asymptotically flat and, therefore, the issue of choosing suitable boundary conditions can be troublesome. Indeed,  problem of which boundary conditions one should impose to compute well-defined QNMs in pure de Sitter space has been subject to several discussions in the literature \cite{Brady, Abdalla, Du,Myung}. Likewise, the problem of adopting suitable boundary conditions for QNMs in anti-de Sitter spacetimes has also been addressed elsewhere \cite{Avis, Peter, Burgess}. In the upcoming section we intend to add to the existing discussion available at the literature.

\section{Boundary Conditions}\label{Sec.Boundary Cond}

Quasinormal modes are solutions of wave equations satisfying specific boundary conditions, generally forming a discrete set. Therefore, the boundary conditions for the fields are a central piece of information behind the quasinormal frequencies \cite{Berti09,Kokkotas,Nollert99,Natario}. The aim of the present section is to discuss the suitable boundary conditions for the quasinormal modes in the class of generalized Nariai spacetimes considered in this article.

\begin{figure}[ht!]
  \centering
  \includegraphics[width=8cm]{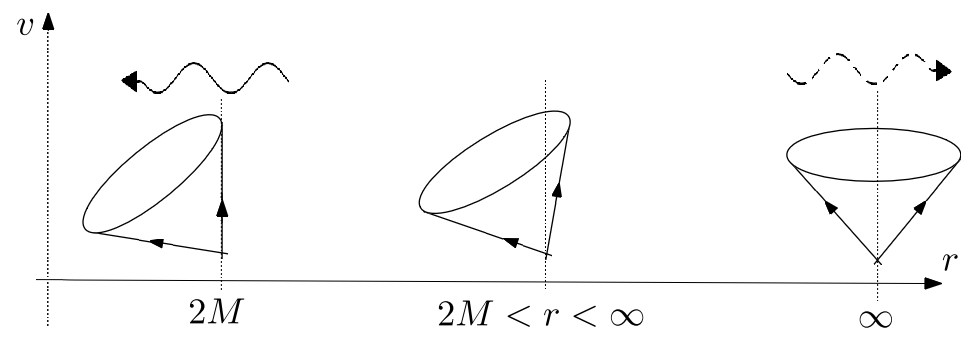}
  \caption{Illustration of the light cone structure in Schwarzschild spacetime, with $r=2M$ being the event horizon. The wavy arrows represent the natural boundary conditions. }\label{FigCones_3}
\end{figure}

In order to motivate the boundary conditions considered in what follows, let us first recall the physical reasoning behind the boundary conditions for the quasinormal modes in Schwarzschild spacetime. Looking at the light cone structure of Schwarzschild spacetime, shown in Fig. \ref{FigCones_3}, we note that at the event horizon ($r=2M$) it is impossible for an observer to increase its radial coordinate, it will inexorably fall towards a smaller value of $r$. Therefore, it is natural to use as boundary conditions at $r=2M$ that the waves are ingoing, which is represented by a infalling wavy arrow in Fig. \ref{FigCones_3}. In its turn, at the infinity, the usual boundary condition is that no wave comes from infinity, whereas some wave can arrive at infinity after scattering by the black hole \cite{Nollert92,Choudhury}. Therefore, at infinity, it is natural to impose that waves are outgoing, as represented by a dashed wavy arrow in Fig. \ref{FigCones_3}.

\begin{figure}[ht!]
  \centering
  \includegraphics[width=8cm]{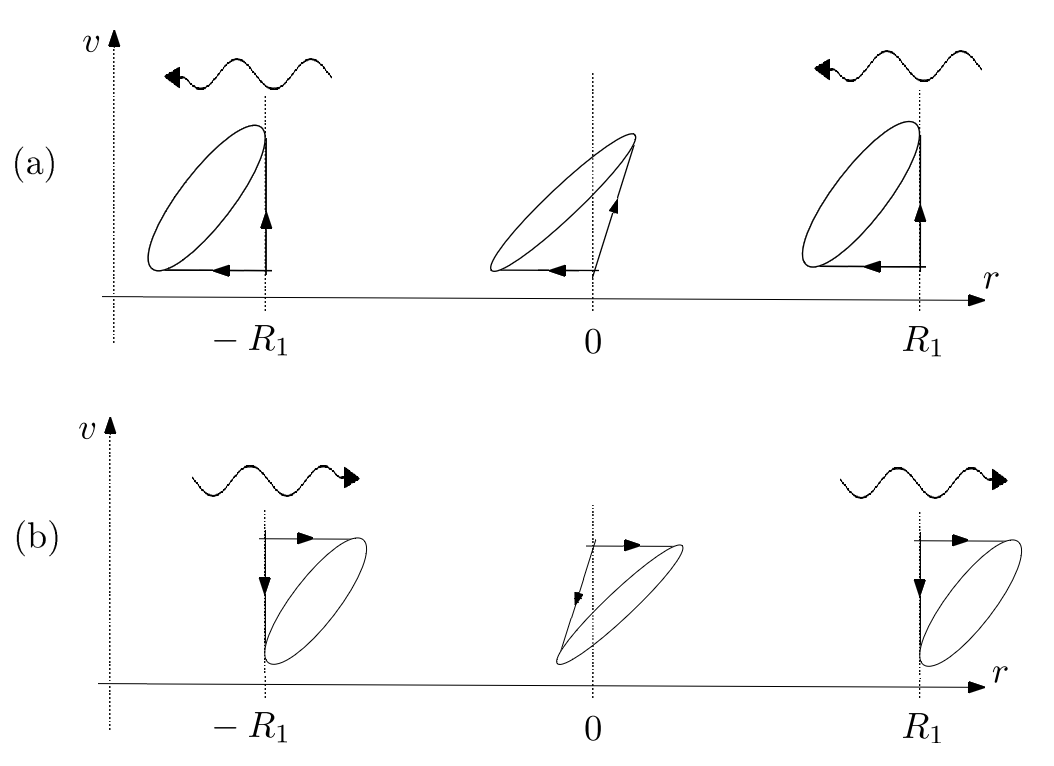}
  \caption{Illustration of the light cone structure of generalized Nariai spacetime. The wavy arrows represent the natural boundary conditions. In the part (a) it is assumed that the time-like vector field $\partial_v$ points to the future, while in part (b) it is assumed that $\partial_v$ points to the past. }\label{FigCones_4}
\end{figure}

Analogously, in order to guess the meaningful boundary conditions for the quasinormal modes in generalized Nariai spacetime, we should look at its light cone structure. Such spacetime is the direct product of the two-dimensional de Sitter spacetime, $dS_2$, with several spheres. In the case of radial wave propagation, namely  $d\theta_j = d\phi_j = 0$, the line element is given by the $dS_2$ line element,
\begin{equation}\label{dS2_LineElem}
  ds^2 = -\,f\, dt^2 + \frac{1}{f}\, dr^2\;,\;\; \textrm{where} \;\; f = 1 -  \frac{r^2}{R_1^2} \;,
\end{equation}
with $R_1$ being a positive constant. Actually, the latter line element represents just a patch of the whole $dS_2$ spacetime, which is rigorously defined as the surface
\begin{equation}\label{Surface_dS}
  - T^2 + X^2 + Y^2 = R_1^2
\end{equation}
immersed into the flat space with Lorentzian line element $ds^2 = - dT^2 + dX^2 + dY^2$. A parametrization of this surface is given by the coordinates $\{t,r\}$ defined by
\begin{equation*}
  \left\{
     \begin{array}{ll}
       T = \sqrt{R_1^2 - r^2} \,\sinh(t/R_1) \\
       X = \sqrt{R_1^2 - r^2} \,\cosh(t/R_1) \\
       Y = r
     \end{array}
   \right.
\end{equation*}
In terms of the parameters $\{t,r\}$, the metric of the surface (\ref{Surface_dS}) is the one given in Eq. (\ref{dS2_LineElem}). In order for the coordinates $T$ and $X$ be real, we must have $r\in [-R_1,R_1]$. Thus, in particular, we find that $-R_1\leq Y \leq R_1$ and $X\geq 0$, so that the surface that defines $dS_2$ is not fully covered by the coordinate system $\{t,r\}$. In addition, note that we should not ignore the negative values of $r$, since this part of the domain of $r$ describes a portion of $dS_2$ that is different from the one covered by $r>0$ \cite{Anninos, Strominger}. This is an important point that differs from what happens in higher-dimensional de Sitter spacetimes \footnote{For instance, $dS_3$ is the surface $- T^2 + X^2 + Y^2 + Z^2 = R^2$ immersed into the flat space $ds^2 = - dT^2 + dX^2 + dY^2 +  dZ^2$. The coordinates $\{t,r,\phi\}$ defined by $T = \sqrt{R_1^2 - r^2}\sinh(t/R_1)$, $X = \sqrt{R_1^2 - r^2}\cosh(t/R_1)$, $Y=r\cos(\phi)$ and $Z=r \sin(\phi)$ cover part of $dS_3$. In this case, note that if we adopt the domain $\phi\in[0,2\pi]$ we just need to consider the positive brunch of $r$.}. Aiming the study of the light cones in $dS_2$, it is useful to introduce the coordinate $v$ defined by the relation $dv = dt + \frac{1}{f}dr$, in terms of which the line element reads
\begin{equation*}
  ds^2 = - f dv^2 + 2 dv dr \,.
\end{equation*}
In particular, since $f> 0$ in the domain $r\in (-R_1,R_1)$, we see that $\partial_v$ is a time-like vector field, so that $v$ can be pictured as a time coordinate. We can assume that this coordinate increases as time passes by, namely that $\partial_v$ points to the future. The null rays of this spacetime are given by
\begin{equation*}
  ds^2 = 0 \; \Rightarrow \; \left\{
                               \begin{array}{ll}
                                 dv = 0 \,, \\
                                 dv = (2/f)dr \,.
                               \end{array}
                             \right.
\end{equation*}
The first light ray, defined by $dv=0$, is tangent to the vector field $\partial_r$. Since the inner product of $\partial_r$ and $\partial_v$ is positive, it follows that the light-like vector field $\partial_r$ points to the past or, in other words, $-\partial_r$ points to the future. Thus, as times passes by, this light ray must decrease its radial coordinate, as illustrated by the horizontal arrows in the line cones in the part (a) of Fig. \ref{FigCones_4}.
The second light ray, given by $dv = (2/f)dr$, is tangent to $2\partial_v + f\partial_r$, which is a null vector field pointing to the future. Since the coefficient in front of $\partial_v$ is positive, it follows that, as time passes by, this light ray increases its coordinate $v$, just as illustrated by the arrows in the line cones in the part (a) of Fig. \ref{FigCones_4}. Also, note that at the boundaries $r=\pm R_1$ we have $f=0$, so that the second light ray points in the direction of $\partial_v$. Then, analysing the light cone structure shown in part (a) of Fig. \ref{FigCones_4}, we can see that an observer cannot increase its radial coordinate when it is at the boundaries $r=\pm R_1$. This, in its turn, suggest that the natural boundary condition for the waves in this spacetime is that they are infalling at both boundaries, as depicted by the wavy arrows. The latter conclusion was based on the arbitrary assumption that $\partial_v$ is oriented to the future. Should we have considered that $\partial_v$ pointed to the past, we would have found the light cone structure depicted at part (b) of Fig. \ref{FigCones_4}. In the latter case, the natural boundary condition is that the waves should be outgoing at both boundaries, as illustrated by the wavy arrows. Due to the symmetry $t\rightarrow -t$ and $r\rightarrow -r$ of the line element (\ref{dS2_LineElem}), it follows that both choices of time orientation for $\partial_v$ are equally valid, there is no preferred choice.

\begin{figure}[ht!]
  \centering
  \includegraphics[width=6cm]{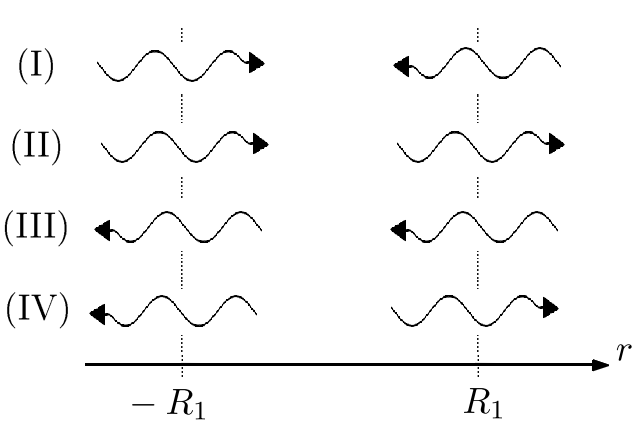}
  \caption{Types of boundary conditions considered in this article. Mathematically, wavy arrows pointing to the right represent $e^{-i\omega(t-r_\star)}$, while wavy arrows pointing to the left represent $e^{-i\omega(t+r_\star)}$.}\label{FigBoundCond}
\end{figure}

Thus, we can say that the natural boundary condition for the waves is that either the waves are infalling at both boundaries or the waves are outgoing at both boundaries. Nevertheless, as we shall see in the sequel,  it turns out that these boundary conditions, although physically motivated, do not lead to quasinormal modes. On the other hand, when we impose that the wave is infalling at one boundary and outgoing at the other boundary, we find what we are looking for: a set discret quasinormal modes. Therefore, in order to our calculations be more complete, in the following sections we will consider four different types of boundaries conditions, the ones defined at Fig. \ref{FigBoundCond}.
In this figure, wavy arrows pointing to the right represent waves moving toward higher values of $r$, mathematically represented by $e^{-i\omega(t-r_\star)}$, while wavy arrows pointing to the left represent waves moving toward lower values of $r$, mathematically represented by $e^{-i\omega(t+r_\star)}$, where the coordinate $r_\star$ will be defined below. As argued in the previous paragraph, boundary conditions (II) and (III) are the ones physically motivated, although they will not lead to quasinormal modes. In contrast, we will see that conditions (I) and (IV) are associated to quasinormal modes.

\section{Integrating the Potential}\label{Sec.Potential}

As we shall see, one important property of studying perturbation equations in the background considered here, the higher-dimensional generalization of the Nariai spacetime presented in Ref. \cite{Carlos-Nariai}, is that all equations turn out to be analytically integrable. Indeed, we are going to see that the problem of solving the perturbation equation for the scalar field (spin 0), the Dirac field (spin $1/2$), and the Maxwell field (spin 1) boils down to integrating the Schr\"{o}dinger-like equation
\begin{equation}\label{SchrodingerEq}
\left[ \frac{d^{2}}{dr_{\star}^{2}} \,+\, \omega^{2}\,-\,V(r_{\star})  \right ] H(r_{\star})= 0 \,,
\end{equation}
with the potential $V(r_\star)$ being given by
\begin{equation}\label{Potential_Generic}
  V(r_{\star}) = A + B \tanh(\gamma \,r_{\star}) + \frac{C}{\cosh^2(\gamma\, r_{\star})} \,,
\end{equation}
where $A$, $B$, $C$ and $\gamma$ are constants with $\gamma > 0$. These constants assume different values depending on the type of the perturbed field. Such $V(r_{\star})$ is contained in the Morse class of integrable potentials, with the case $A=B=0$ being the well-known P\"{o}schl-Teller potential, see \cite{Khare}. In order to solve the latter ordinary differential equation, let us define a new independent variable defined by
\begin{equation*}
  y = \frac{1}{2} + \frac{1}{2} \tanh(\gamma \, r_\star).
\end{equation*}
Assuming that the domain of $r_\star$ is the entire real line, we find that $y\in(0,1)$, with the boundaries $r_\star = \pm\infty$ being given by $y = 0$ and $y=1$.  In particular, near the boundaries, the relation between the coordinates $r_\star$ and $y$ assumes the simpler form
\begin{equation}\label{yr_boudary}
\left\{
  \begin{array}{ll}
   r_\star \rightarrow -\infty \; \Rightarrow \; y \simeq e^{2\gamma\, r_\star} \,, \\
    r_\star \rightarrow +\infty \; \Rightarrow \; (1-y) \simeq   e^{-2\gamma\, r_\star} \,.
  \end{array}
\right.
\end{equation}

Now, let us define the constant parameters $a$, $b$, and $c$ as follows
\begin{widetext}
\begin{equation}\label{abc}
     \left\{  \begin{array}{ll}
       a = \frac{1}{2\gamma}\left(\gamma  +  \sqrt{A - B - \omega^2} -  \sqrt{A +  B - \omega^2} + \sqrt{\gamma^2 - 4 C} \right) \,, \\
       b = \frac{1}{2\gamma}\left( \gamma  + \sqrt{A - B - \omega^2} -  \sqrt{A +  B - \omega^2} - \sqrt{\gamma^2 - 4 C}   \right)  \,,\\
       c =   \frac{1}{\gamma}\sqrt{A - B - \omega^2} + 1 \,,
     \end{array}  \right.
\end{equation}
and, instead of $H(r_\star)$, let us use the dependent variable $G(y)$ defined by
\begin{equation}\label{fieldH}
  H(r_{\star}) = y^{(c-1)/2}\,( 1- y )^{\frac{1}{2}(a+b-c)} \,G(y) \,.
\end{equation}
Then, after some algebra, one can check that the function $G(y)$ obeys the equation
\begin{equation}\label{HeperGeomEq}
  y(1-y) \frac{d^2G}{dy^2} + \left[ c - y(a+b+1)\right]\frac{d G}{dy }  - a b \,G = 0 .
\end{equation}
This is the hypergeometric equation, whose general solution is given by
\begin{equation}\label{G(y)}
  G(y) = \alpha \,F(a,b,c;y) + \beta \,y^{(1-c)}\,F(1+a-c,1+b-c,2-c;y) \,,
\end{equation}
where $F$ is the hypergeometric function (usually denoted by $_2F_1$), while $\alpha$ and $\beta$ are arbitrary integration constants that can be fixed by the boundary conditions. Summing up these results, we conclude, from Eqs. (\ref{fieldH}) and (\ref{G(y)}), that the solution for the function $H$ obeying Eq. (\ref{SchrodingerEq}) is given by
\begin{equation}\label{SolutionH}
  H = ( 1 - y )^{\frac{1}{2}(a+b-c)} \left[\,\alpha\, y^{(c-1)/2}\, F(a,b,c;y) + \beta \,y^{-(c-1)/2}\,F(1+a-c,1+b-c,2-c;y)  \, \right] \,.
\end{equation}
Since $F(a,b,c;0)=1$, it turns out that the latter way of writing the solution is particularly useful to apply the boundary conditions at $y=0$, i.e. $r_\star=-\infty$. Indeed, using Eq. (\ref{yr_boudary}), one can promptly verify that the following limit holds
\begin{equation}\label{SolutionH_y=0}
   \left. H\right|_{r_\star\rightarrow -\infty} = \,\alpha\, e^{\gamma(c-1)r_\star}\,  + \beta \,e^{-\gamma(c-1)r_\star}   \,,
\end{equation}
which will be of relevance to impose the boundary conditions at $r_\star=-\infty$. On the other hand, in order to apply the boundary conditions at $y=1$, i.e. $r_\star=+\infty$, it is more useful to write the hypergeometric functions as functions of $(1-y)$, so that they become unit at the boundary. This can be done rewriting the hypergeometric functions appearing in Eq. (\ref{SolutionH}) by means of the following identity \cite{Abramowitz}:
\begin{align}
  F(a,b,c;y) \,=\,&  \frac{\Gamma(c-a-b)\Gamma(c)}{ \Gamma(c-a) \Gamma(c-b) }  F(a,b,a+b-c+1;1-y) \nonumber\\
 &+ \frac{\Gamma(a+b-c)\Gamma(c)}{ \Gamma(a) \Gamma(b) }  (1-y)^{(c-a-b)}F(c-a,c-b,c-a-b+1;1-y)\,, \label{HyperGeomIdentity}
\end{align}
where $\Gamma$ stands for the gamma function. Doing so, and using Eq. (\ref{yr_boudary}) we eventually arrive at the following behaviour  of the solution at
$r_\star=+\infty$:
\begin{align}
  \left. H\right|_{r_\star\rightarrow +\infty} \,\simeq & \,    e^{-\gamma(a+b-c)r_\star} \left[ \,\alpha\, \frac{\Gamma(c-a-b)\Gamma(c)}{ \Gamma(c-a) \Gamma(c-b)} +
\beta \,   \frac{\Gamma(c-a-b)\Gamma(2-c)}{ \Gamma(1-a) \Gamma(1-b)   } \,\right] \nonumber\\
& + e^{\gamma(a+b-c)r_\star}  \left[\,
  \alpha\, \frac{\Gamma(a+b-c)\Gamma(c)}{ \Gamma(a) \Gamma(b) }  + \beta \, \frac{\Gamma(a+b-c)\Gamma(2-c)}{ \Gamma(a-c+1) \Gamma(b-c+1) }    \,\right] \,. \label{SolutionH_y=1}
\end{align}
\end{widetext}

\section{Scalar perturbations}\label{Sec.Scalar}

Now, with the integration of the general equation (\ref{SchrodingerEq}) at hand, we are ready to move on and study the perturbation of several matter fields. Let us start with the perturbations in a scalar field $\Phi$ of mass $\mu$. For the study of the quasinormal modes of scalar fields in other backgrounds, see \cite{Scalar1,Scalar2,Scalar3,Scalar4}.

The equation obeyed by the scalar field while it propagates in the background (\ref{nariai-metric}) is the Klein-Gordon equation given by
\begin{equation}\label{KGE}
\frac{1}{\sqrt{\left | g^{0} \right |}} \,\partial_{\mu} \left( g^{0\,\mu\nu}\sqrt{ \left | g^{0} \right |} \,\partial_{\nu} \right)\Phi \,=\,\mu^{2}\Phi \,.
\end{equation}
 In order to accomplish the integrability of this equation, it is useful to introduce the tortoise coordinate $r_{\star}$ defined by the equation
\begin{equation}\label{TC}
dr_{\star}=\frac{1}{f(r)}\,dr \;\;\Rightarrow \;\; r_{\star}=R_{1} \, \text{arctanh}\left(\frac{r}{R_{1}}\right) \,.
\end{equation}
In particular, note that the tortoise coordinate maps the domain between two horizons, $r\in(-R_1,R_1)$, into the interval $r_{\star} \in (-\infty, \infty)$. In terms of this coordinate, the line element is written as
\begin{equation}\label{FF1}
ds^{2}=\frac{1}{\cosh^{2}( r_{\star}/R_{1})}\,(-dt^{2} + dr_{\star}^{2}) + \sum_{j=2}^{d}R_{j}^{2}\,d\Omega_{j}^{2} \,.
\end{equation}
Thus, writing  \eqref{KGE} in these coordinates, we eventually arrive at the following field equation
\begin{equation*}\label{KGE1}
\left[ \cosh^{2}( r_{\star}/R_{1})  (\partial_{r_{\star}}^{2}-\partial_{t}^{2}) +\sum_{j=2}^{d}\frac{\Delta_{j}}{R_{j}^2} -\mu^{2} \right ]\Phi =0 \,,
\end{equation*}
where
\begin{equation}
\Delta_{j} \equiv \frac{1}{\text{sin}\theta_{j}} \partial_{\theta_{j}}(\text{sin}\theta_{j}\,\partial_{\theta_{j}}) \,+\,\frac{1}{\text{sin}^{2}\theta_{j}} \partial^{2}_{\phi_{j}}
\end{equation}
is the Laplace-Beltrami operator on the unit sphere.  The eigenfunctions of $\Delta_{j}$ are the well-known scalar spherical harmonics, $Y_{\ell_{j}}^{m_{j}}(\theta_{j}, \phi_{j})$, with eigenvalues determined by the equation
\begin{equation}
\Delta_{j}Y_{\ell_{j}}^{m_{j}}(\theta_{j}, \phi_{j})\,=\,-\ell_j(\ell_j \,+\, 1)\,Y_{\ell_{j}}^{m_{j}}(\theta_{j}, \phi_{j}) \,,
\end{equation}
with $\ell_j$ and $m_j$ being integers, $|m_j|\leq \ell_j$, and $\ell_j \geq 0$.
Thus, it is fruitful to expand the scalar field in terms of the spherical harmonics as follows
\begin{equation}\label{ExpansionSacalar}
\Phi=\int d\omega \sum_{\ell, m}\phi^{\omega}_{\ell,m}(r_{\star}) e^{-i\omega t}\,\mathcal{Y}_{\ell,m} ,
\end{equation}
where
\begin{equation}\label{YLM}
  \mathcal{Y}_{\ell,m} = \prod_{j=2}^d \, Y_{\ell_j}^{m_j}(\theta_j,\phi_j)  \,,
\end{equation}
At Eq. (\ref{ExpansionSacalar}) we have taken into account the fact that $t$ is a cyclic coordinate of the metric, so that it is useful to decompose the temporal dependence of the field $\Phi$ in the Fourier basis. The sum over the collective index $\{\ell, m \}$ means that we are summing over all values of the set $\{\ell_{2}, m_{2}, \ell_{3}, m_{3}, \ldots , \ell_{d}, m_{d} \}$.

Then, inserting the expansion \eqref{ExpansionSacalar} into the field equation lead us to the following ordinary differential equation for the components $\phi^{\omega}_{\ell,m}$:
\begin{equation}\label{WE}
\left[ \frac{d^{2}}{dr_{\star}^{2}} \,+\, \omega^{2}\,-\,V(r_{\star})  \right ]\phi^{\omega}_{\ell,m} \,=\,0 \,,
\end{equation}
where the potential $V(r_\star)$ is the one studied in the previous section, see Eq. (\ref{Potential_Generic}), with the parameters $A$, $B$, $C$, and $\gamma$ given by:
\begin{equation*}
  A = 0 ,\; B = 0,\; C = \mu^{2} +\sum_{j=2}^d \frac{\ell_{j}(\ell_j + 1)}{R_j^2} ,\; \gamma = 1/R_1\,.
\end{equation*}
Inserting these parameters into Eq. (\ref{abc}), we find that the constants appearing at the hypergeometric equation are given by
\begin{equation}\label{abcScalar}
      \begin{array}{ll}
      a = \frac{1}{2} + i\sqrt{\mu^2 R_1^2 +\left[\sum_{j=2}^d  \frac{\ell_{j}(\ell_j + 1)R_1^2}{R_j^2} \right] - 1/4 }  ,   \\
      b = \frac{1}{2} - i\sqrt{\mu^2 R_1^2 +\left[\sum_{j=2}^d  \frac{\ell_{j}(\ell_j + 1)R_1^2}{R_j^2} \right] - 1/4 } ,  \\
      c = 1+ i\,R_1\,\omega \,.
    \end{array}
  \end{equation}

Now, we are ready to impose the boundary conditions. Let us start with the boundary condition I, described at Fig. \ref{FigBoundCond}. In this case, the field is assumed to move to increasing $r_\star$ at the boundary $r_\star = -\infty$  while at the boundary $r_\star = +\infty$ it should move towards lower values of $r_\star$. Since the time dependence of the mode $\phi^{\omega}_{\ell,m}$ is of the type $e^{-i\omega t}$, this means that  $\phi^{\omega}_{\ell,m}$ should behave as $e^{i\omega r_\star}$ at  $r_\star = -\infty$, while it should go as $e^{-i\omega r_\star}$ at $r_\star = +\infty$. Noting that, in the case considered in this section, Eq. (\ref{SolutionH_y=0}) translates to
\begin{equation}\label{Scalar-Iy=0}
   \left. \phi^{\omega}_{\ell,m}\right|_{r_\star\rightarrow -\infty} = \,\alpha\, e^{i \omega r_\star}\,  + \beta \,e^{-i \omega r_\star}   \,.
\end{equation}
Thus, for the boundary condition I to hold we need to set $\beta=0$. In such a case, the behaviour of the scalar mode at the boundary $r_\star = +\infty$ is given by
\begin{align}\label{Scalar-Iy=1}
  \left. \phi^{\omega}_{\ell,m}\right|_{r_\star\rightarrow +\infty}  \simeq \,&   \alpha \,   \frac{\Gamma(c-a-b)\Gamma(c)}{ \Gamma(c-a) \Gamma(c-b)}
e^{i \omega r_\star}   \nonumber \\
&  + \alpha  \, \frac{\Gamma(a+b-c)\Gamma(c)}{ \Gamma(a) \Gamma(b) } e^{-i \omega r_\star}  \,,
\end{align}
a relation that stems from Eqs. (\ref{SolutionH_y=1}) and (\ref{abcScalar}). The boundary condition I imposes that the coefficient multiplying $e^{i \omega r_\star}$ should vanish. Since $\alpha$ cannot be zero, otherwise the mode would vanish identically, we need the combination of the gamma functions to be zero. Now, once the gamma function has no zeros, the way to achieve this is to let the gamma functions at the denominator to diverge, $\Gamma(c-a)=\infty$ or $\Gamma(c-b)=\infty$. Since the gamma function diverge only at non-positive integers, we are led to the following constraint:
\begin{equation*}
  c - a = -n \;\; \textrm{or} \;\; c - b = -n\,,\;\; \textrm{where }   n\in\{0,1,2,\cdots\}.
\end{equation*}
Using Eq. (\ref{abcScalar}), we find that these constraints translate to
\begin{equation*}
  \omega = \pm\sqrt{\mu^2  + \sum_{j=2}^d  \frac{\ell_{j}(\ell_j + 1)}{R_j^2}   - \frac{1}{4R_1^2} } + \frac{i}{2R_1} (2n+1),
\end{equation*}
where $n$ is any non-negative integer. These frequencies are the only ones compatible with the boundary condition I, they are the so-called frequencies of the quasinormal modes. Note the presence of the imaginary part in the frequency, which accounts for a damping on the field, a feature of perturbations in the presence of horizons. Moreover, note that the square root could also lead to an imaginary part of the frequency, in which case the perturbation mode would be solely dumped, with no characteristic oscillation. For instance, the case of a massless field the frequency spherically symmetric mode ($\ell_j=0$) will be purely imaginary.

Now, let us investigate the boundary condition II. In this case the mode $\phi^{\omega}_{\ell,m}$ should behave as $e^{i\omega r_\star}$ at both boundaries
$r_\star = \pm\infty$, as depicted at Fig. \ref{FigBoundCond}. Thus, since the behaviour at $r_\star = -\infty$ is the same as at boundary condition I, it follows that Eq. (\ref{Scalar-Iy=1}) remains valid for the boundary condition II. The only difference is that at Eq. (\ref{Scalar-Iy=1}) we should eliminate the term $e^{-i\omega r_\star}$, which is possible only if either $\Gamma(a)$ or $\Gamma(b)$ diverge. Since $a$ and $b$ do not depend on the frequency $\omega$, see Eq. (\ref{abcScalar}), it follows that the constraints $a=-n$ and $b=-n$, with $n$ a non-negative integer, would represent restriction on parameters that are already fixed, like the mass $\mu$ and the radii $R_j$ that describe the background. Therefore, we conclude that, generally, we have no solution for the perturbation when the boundary condition II is assumed.

For the boundary condition $III$, the mode $\phi^{\omega}_{\ell,m}$ should behave as $e^{-i\omega r_\star}$ at both boundaries $r_\star = \pm\infty$.
Therefore, at  Eq. (\ref{Scalar-Iy=0}) we should set $\alpha=0$, in which case we are left with the following form at $r_\star = +\infty$:
\begin{align}
  \left. \phi^{\omega}_{\ell,m}\right|_{r_\star\rightarrow \infty}& \simeq      \beta   e^{i \omega r_\star}
  \frac{\Gamma(c-a-b)\Gamma(2-c)}{ \Gamma(1-a) \Gamma(1-b)   }  \nonumber\\
& +  \beta  e^{-i \omega r_\star}
   \frac{\Gamma(a+b-c)\Gamma(2-c)}{ \Gamma(a-c+1) \Gamma(b-c+1) }  . \label{SolutionH_y=1III}
\end{align}
In order to eliminate the term  $e^{i\omega r_\star}$, we need to set $1-a = -n$ or $1-b=-n$, with $n\in\{0,1,2,\cdots\}$. Just as in the case of boundary II, this constraint cannot me satisfied in general. Thus, we have no quasinormal modes obeying the boundary condition III.

Finally, for the boundary condition IV, the field must behave as $e^{-i\omega r_\star}$ at $r_\star = -\infty$ while at  $r_\star = +\infty$ it should go as
$e^{i\omega r_\star}$. Hence, at Eq. (\ref{SolutionH_y=1III}) we should get rid of the term $e^{-i\omega r_\star}$, which can be accomplished by setting
$a-c+1=-n$ or $b-c+1=-n$, with $n$ being a non-negative integer. The latter constraints along with Eq. (\ref{abcScalar}) lead to the following quasinormal frequencies:
\begin{equation*}
  \omega = \pm\sqrt{\mu^2  + \sum_{j=2}^d  \frac{\ell_{j}(\ell_j + 1)}{R_j^2}   - \frac{1}{4R_1^2} } - \frac{i}{2R_1} (2n+1).
\end{equation*}
This spectrum is almost equal to the one found for the boundary condition I, the only difference being the sign of the imaginary part. Thus, while for the boundary condition I the modes dwindle for $t\rightarrow-\infty$ and diverge for $t\rightarrow+\infty$, for the boundary condition IV it is the other way around.

\section{Maxwell Field}\label{Sec.Maxwell}

In this section we shall consider the perturbations on the Maxwell field $\bl{\mathcal{A}}$, a massless spin one field. In this case we shall assume that the electromagnetic charges of the background are zero, namely $Q_1=Q_j=0$, so that we have a vanishing Maxwell field in the background, $\bl{\mathcal{A}}^0=0$. This is important to validate the separability of the perturbations in the background metric and the matter fields, as discussed at Sec. \ref{Sec.Problem}. In particular this means that the radii $R_1$ and $R_j$ are all equal in such a case, see Eq. (\ref{radii}). For the calculation of quasinormal modes of spin 1 fields in other backgrounds, see \cite{Maxwell1,Maxwell2,Maxwell3,Maxwell4,Maxwell4}.

The source-free  Maxwell field equation is given by
\begin{equation}\label{Maxwell-equation}
\nabla_{\mu}F^{\mu\nu}\,=\,0 \,, \quad \text{with} \quad F_{\mu\nu}\,=\,\partial_{\mu}\mathcal{A}_{\nu}\,-\,\partial_{\nu}\mathcal{A}_{\mu}\,.
\end{equation}
Taking into account the symmetries of the background considered here, a suitable ansatz to the gauge field in order to accomplish separability of the field equation is provided by
\begin{align*}
 \bl{\mathcal{A}} = &  e^{-i\omega t} \bigg[ \,\tilde{A}^0\,\mathcal{Y}_{\ell,m} \,dt + \tilde{A}^1\,\mathcal{Y}_{\ell,m} \,dr_\star   \\
 &  - \sum_{j=2}^{d} \tilde{A}^j   \, \left( \frac{\partial_{\phi_j} \mathcal{Y}_{\ell,m} }{\sin\theta_j} \,d\theta_j
  -     \sin\theta_j \,\partial_{\theta_j} \mathcal{Y}_{\ell,m}\, d\phi_j \right) \bigg]\,,
\end{align*}
where $\tilde{A}^0$, $\tilde{A}^1$ and $\tilde{A}^j$ are arbitrary functions of $r_\star$, while $\mathcal{Y}_{\ell,m}$ has been defined in Eq. (\ref{YLM}).   This way of writing the degrees of freedom of the Maxwell field comes from the fact that this field has spin one field and, therefore, the spherical symmetries should show up it terms of vector spherical harmonics, as explained in detail at the appendix \ref{Appendix.Maxwell}.

Now, inserting this ansatz into Maxwell's source-free equation, $\nabla^\mu F_{\mu\nu}=0$, we can find the differential equations obeyed by the components $\tilde{A}^0$, $\tilde{A}^1$ and $\tilde{A}^j$. In order to accomplish this, it is useful to use the function $\check{A}^1$ defined by
\begin{equation}
\check{A}^1 = \cosh^2(r_\star/R_1)\,\left[ \frac{d}{dr_\star}\tilde{A}^0(r_\star) + i \omega \tilde{A}^1(r_\star) \right] \,,
\end{equation}
instead of the degree of freedom $\tilde{A}^1$. Doing so, it follows from the identity $\partial_t (\nabla^\mu F_{\mu r_\star}) - \partial_{r_\star} (\nabla^\mu F_{\mu t}) = 0$, which is a consequence of the components $\nabla^\mu F_{\mu t}=0$ and $\nabla^\mu F_{\mu r_\star}=0$ of Maxwell's field equation, that $\check{A}^1$ obeys the following Schr\"{o}dinger equation:
 \begin{equation}\label{WaveEqMaxwell}
   \left[\frac{d^2 }{dr_\star^2} +  \omega^2 - \sum_{j=2}^{n} \,\frac{\ell_j(\ell_j + 1)}{R_j^2\,\cosh^2(r_\star/R_1)}\,\right]  \,\check{A}^1 = 0 \,.
\end{equation}
This is the same equation obeyed by the scalar field mode $\phi^{\omega}_{\ell m}$ when the scalar field has vanishing mass ($\mu=0$). Thus, the quasinormal spectrum associated to this component of the Maxwell field must be the same of the massless scalar field.
Then, assuming that $\check{A}^1$ is a solution of Eq. (\ref{WaveEqMaxwell}), the identity $\nabla^\mu F_{\mu t}=0$ lead to the fact that $\tilde{A}^0$ is related to $\check{A}^1$ by the following equation:
\begin{equation}\label{A0}
 \tilde{A}^0 =
 \left(\sum_{j=2}^{n} \, \frac{\ell_j(\ell_j + 1)}{R_j^2}\right)^{-1}\, \frac{d}{dr_\star}\check{A}^1 \,.
\end{equation}
Thus, the component $\tilde{A}^0$ of the gauge field must have the same spectrum of $\check{A}^1$. Finally, imposing equations $\nabla^\mu F_{\mu \theta_j}=0$, while assuming that (\ref{WaveEqMaxwell}) and (\ref{A0}) hold, we conclude that $\tilde{A}^j$ must obey the same differential equation of $\check{A}^1$, namely Eq. (\ref{WaveEqMaxwell}).

Summing up, we have obtained that all the degrees of freedom of the Maxwell field have the same spectrum of the massless scalar field. In particular, this means that for the boundary conditions II and III we have no quasinormal modes, while for the boundary conditions I and IV the frequencies must have the form
\begin{equation*}
  \omega = \pm\sqrt{ \sum_{j=2}^d  \frac{\ell_{j}(\ell_j + 1)}{R_j^2}   - \frac{1}{4R_1^2} } \pm \frac{i}{2R_1} (2n+1).
\end{equation*}

\section{Spinorial Field}\label{Sec.Spinor}

Finally, let us consider perturbations on a spinorial field $\bl{\Psi}$ satisfying the Dirac equation minimally coupled to the electromagnetic field of the background
\begin{equation}\label{Dirac-equation}
\gamma^{\alpha}(\nabla_{\alpha}-iq \mathcal{A}^0_\alpha )\bl{\Psi}\,=\,\mu \bl{\Psi}\,,
\end{equation}
where $\gamma^{\alpha}$ are the Dirac matrices, $q$ and $\mu$ denote, respectively, the electric charge and mass of the spinorial field, while $\bl{\mathcal{A}}^0$ stands for the background Maxwell field. For previous works on quasinormal modes of Dirac fields in other backgrounds, see \cite{Spinor1,Spinor2,Spinor3}.

In $D=2d$ dimensions, the spinorial field has $2^d$ degrees of freedom, which can be written in terms of the column matrices
\begin{equation*}
  \xi^{+} = \left[
              \begin{array}{c}
                1 \\
                0 \\
              \end{array}
            \right] \;\; \textrm{and} \;\;
 \xi^{-} = \left[
              \begin{array}{c}
                0 \\
                1 \\
              \end{array}
            \right]
\end{equation*}
as follows
\begin{equation}\label{spinor}
\boldsymbol{\Psi} \,=\,\sum_{\{s\}}\Psi^{s_{1}s_{2}\dots s_{d}}\xi^{s_{1}}\otimes \xi^{s_{2}} \otimes \dots \otimes \xi^{s_{d}}\,,
\end{equation}
where the indices $s_a$ run over $\{-,+\}$. In order to solve the Dirac equation, we need to separate the degrees of freedom of the field, which can be quite challenging in general. Nonetheless, we should note that the spacetime considered here  is the direct product of two-dimensional spaces, which is exactly the class of spaces studied in Ref. \cite{Joas}. Indeed, the main goal of this reference is to show that the Dirac equation is separable in such backgrounds.

Following the procedure of  Ref. \cite{Joas}, we must introduce an orthonormal frame of vector fields, which in the case of our background is given by
\begin{align}
  \bl{e}_1  &= - i \cosh(r_\star/R_1) \partial_t \;,\;\;   \bl{e}_{\tilde{1}} =  \cosh(r_\star/R_1) \partial_{r_\star} \;, \nonumber\\
    \bl{e}_j  &=  \frac{1}{R_{j}\,\text{sin}\theta_{j}} \partial_{\phi_j} \;,\;\;
\quad\quad \; \bl{e}_{\tilde{j}}  =  \frac{1}{R_{j}} \partial_{\theta_j} \;. \label{BCF}
\end{align}
Associated with this frame we have the spin coefficients $\omega_{\alpha\beta\gamma}$ defined by
\begin{equation}
\nabla_{\alpha}\boldsymbol{e}_{\beta} \,=\,\omega_{\alpha\beta}^{\phantom{\alpha\beta}\gamma}\boldsymbol{e}_{\gamma} \,.
\end{equation}
Then, assuming the decomposition of the spinorial field as
\begin{equation*}
\Psi^{s_{1}s_{2}\dots s_{d}} = \Psi_{1}^{s_{1}}(t, r_\star)\Psi_{2}^{s_{2}}(\phi_{2},\theta_{2})\ldots \Psi_{d}^{s_{d}}(\phi_{d}, \theta_{d}) ,
\end{equation*}
it follows from Ref.  \cite{Joas} that the component $\Psi_{1}^{s_{1}}(t, r_\star)$ obeys the following differential equation
\begin{align}
  \bigg[   \partial_{\tilde{1}} + & \frac{\omega_{1\tilde{1}1}}{2}  -i q \mathcal{A}^0_{\tilde{1}} - i s_{1} \left(\partial_{1}+
\frac{\omega_{\tilde{1}1\tilde{1}}}{2} - i q \mathcal{A}^0_{1}\right)     \bigg] \Psi_{1}^{s_{1}} \nonumber\\
 & = (L -i s_{1}\,\mu)\,\Psi_{1}^{-s_{1}} \,. \label{DifEqSpinor}
\end{align}
The parameter $L$ appearing in the latter equation is a separation constant that depends on the angular modes. In particular, in the case of vanishing magnetic charges $Q_{j}$, they are determined by the eigenvalues $\lambda_{j}$ of the Dirac operator on the unit sphere $S^{2}$ according to the following relation
\begin{equation}\label{L}
L = \sqrt{\frac{\lambda^{2}_{2}}{R_2^2}+ \frac{\lambda^{2}_{3}}{R_3^2}  +\ldots+\  \frac{\lambda^{2}_{d}}{R_d^2} } \,  ,
\end{equation}
with $ \lambda_{j} = \pm  1, \pm  2,  \ldots$, as proved at Appendix \ref{Appendix.Angular}. These $\lambda_{j}$ are the analogous of the representation labels $\ell_j$ of the spherical harmonics. For the general case, in which the magnetic charges of the background are non-vanishing, the parameters $\lambda_j$, and, hence, the separation constant $L$, must be found numerically.

%
%
%
%
%

The only nonzero spin coefficients of the frame considered here are
\begin{align*}
  \omega_{1\tilde{1}1}&=  -\,\omega_{11\tilde{1}}= - \frac{1}{R_{1}}\,\text{sinh}(r_{\star}/R_1) \,,\\
 \omega_{j\tilde{j}j} &= -\,\omega_{jj\tilde{j}} =   \frac{1}{R_{j}}\,\text{cot}\theta_{j} \,.
\end{align*}
Also, the nonzero components of the background electromagnetic field in the considered frame are given by
\begin{equation*}\label{gauge-field2}
\mathcal{A}^0_{1} = -i Q_{1}R_{1}\text{sinh}(r_{\star}/R_1)  , \;\;
\mathcal{A}^0_{j} =  Q_{j}R_{j}\,\text{cot}\theta_{j} \,.
\end{equation*}
Then, expanding the time dependence of $\Psi_{1}^{s_{1}}$ at the Fourier basis,
\begin{equation}
\Psi_{1}^{s_{1}}(t, r_\star)\,=\, e^{-i\omega t}\psi^{s_{1}}(r_\star) \,,
\end{equation}
it follows that the field equation (\ref{DifEqSpinor}) yields
\begin{align}
 \bigg[    \frac{d}{dr_\star} &  + i s_{1} \omega   + \left( i  s_{1} q  Q_{1}R_{1} -  \frac{1}{2R_{1}}  \right) \text{tanh}(r_{\star}/R_1)  \bigg] \psi_{1}^{s_{1}}    \nonumber \\
 &  = \frac{(L -is_{1}\,\mu )}{\cosh(r_\star/R_1)}\,\psi_{1}^{-s_{1}} \,.  \nonumber
\end{align}
Note that this first order differential equation mixes the components $\psi_{1}^{+}$ and $\psi_{1}^{-}$ of the spinorial field. In order to separate these components we need to differentiate once more this equation, which, after some algebra, lead to the equation
\begin{equation}\label{WED}
\left[  \frac{d^{2}}{dr_\star^{2}} \,+\,\omega^{2} \,-\, V(r_\star)
 \right ]\psi^{s_{1}}\,=\,0 \,,
\end{equation}
where the potential $V(r_\star)$ is the one considered at Eq. (\ref{Potential_Generic}) with the parameters $A$, $B$ and $C$ being given by
\begin{align}
  A &\,=\,\frac{1}{4 R_1^2} -q\,Q_{1} (i s_1 +q\,Q_{1} \, R_1^2 )\,,  \nonumber \\
       B &\,=\, -\frac{\omega}{R_1}\, (is_{1} + 2\,q \,Q_{1} \, R_1^2) \,,  \label{ABC_Spin}\\
       C &\,=\,\mu^{2} + L^2 +  \frac{1}{4 R_1^2} + q^{2}\,Q^{2}_{1}  R_1^2 \,,\nonumber
\end{align}
while the parameter $\gamma$ is given by $1/R_1$. In particular, note that the potential above is complex, whereas in most problems of QNMs the potentials turn out to be real. Although it is possible to make field redefinitions in order to make the potential real \cite{Guven}, we shall not do it here. Moreover, note that the potential does not vanish at $r_\star \rightarrow \pm \infty$, so that the solution at the boundaries is not of the plane-wave type. In order to see the latter fact more clearly, let us calculate the parameters $a$, $b$ and $c$ by plugging (\ref{ABC_Spin}) into (\ref{abc}), which lead us to:
\begin{widetext}
\begin{align}
a &= i R_1 \sqrt{\mu^2 + q^2 Q_1^2 R_1^2 + L^2} + (1+s_1)\left( \frac{1}{4} - i \omega \frac{R_1}{2}\right) -i (1-s_1)\frac{q Q_1 R_1^2}{2} \,,
\nonumber \\
b &= -i R_1 \sqrt{\mu^2 + q^2 Q_1^2 R_1^2 + L^2} + (1+s_1)\left( \frac{1}{4} - i \omega \frac{R_1}{2}\right) -i (1-s_1)\frac{q Q_1 R_1^2}{2} \,, \label{abcSpinor}\\
c &=  \frac{1}{2} + i s_1 \left(q Q_1 R_1^2 - \omega R_1 \right) \,. \nonumber
\end{align}
\end{widetext}

%
%
%
%

Now, with Eq. (\ref{abcSpinor}) at hand, we are ready to impose the boundary conditions in order to investigate the quasi-normal modes. Since all we need, for this end, are the asymptotic behaviour obtained in Eqs. (\ref{SolutionH_y=0}) and (\ref{SolutionH_y=1}), and since they depend just on the exponents $\gamma(c-1)$ and $\gamma(a+b-c)$, it is useful write the explicit expressions for these combinations:
\begin{equation}\label{ExponentsSpinor}
      \begin{array}{ll}
       \gamma (c-1) &= - i s_1 \omega + i s_1 q Q_1 R_1 - \frac{1}{2R_1} \,, \\
\gamma (a+b-c) &= - i  \omega - i  q Q_1 R_1 + s_1\frac{1}{2R_1}  \,. \\
     \end{array}
\end{equation}
Then, for instance, let us impose the boundary condition I for the component $s_1 = +$ of the spinorial field. In this case, Fig. \ref{FigBoundCond} tells us that the field must have a dependence of the type $e^{i\omega r_\star}$ at $r_\star \rightarrow -\infty$, while it must goes as $e^{-i\omega r_\star}$ at $r_\star \rightarrow \infty$. Thus, inserting Eq. (\ref{ExponentsSpinor}), with  $s_1 = +$, into (\ref{SolutionH_y=0}), lead us to conclude that we must set $\alpha=0$. Then, inserting $\alpha=0$ into (\ref{SolutionH_y=1}) and demanding the behaviour $e^{-i\omega r_\star}$ at $r_\star \rightarrow \infty$ yield the constraint $1-a = -n$ or $1-b = -n$, with $n$ being a non-negative integer. These imply that the frequencies must be given by
\begin{equation}\label{SpectrumSpinor1}
  \omega =  \pm\,\sqrt{\mu^2 + q^2 Q_1^2 R_1^2 + L^2} + \frac{i}{2R_1}(2n+1) \,,
\end{equation}
with $n\in\{0,1,2,\cdots\}$. At this point, it is worth recalling that $L$ is a separation constant of the Dirac equation that is related to the angular mode of the field.

Likewise, imposing the boundary condition $I$ to the component $s_1= -$ of the spinorial field, we  find that we must set $\beta = 0$ at Eq. (\ref{SolutionH_y=0}) and then $c-a = -n$ or $c-b = -n$, with $n$ being a non-negative integer. This, in its turn, lead to the same spectrum obtained for the component  $s_1= +$, namely \eqref{SpectrumSpinor1}.

Analogously, imposing the boundary conditions II and III for the spinorial field, we find that no quasi-normal mode exists in these cases, just as happens with the scalar and Maxwell's fields. On the other hand, imposing the boundary condition IV, we find that the quasi-normal frequencies are given by
\begin{equation*}\label{SpectrumSpinor2}
  \omega =  \pm\,\sqrt{\mu^2 + q^2 Q_1^2 R_1^2 + L^2} - \frac{i}{2R_1}(2n+1) \,,
\end{equation*}
where $n\in\{0,1,2,\cdots\}$.

\subsection{Analysing the regularity of the solution}

Note that the solution for the spinorial field equation is not exactly a plane wave at the boundaries, which is a consequence of the fact that the potential $V(r_\star)$ does not vanish at $r_\star \pm \infty$. Indeed, computing the asymptotic form of the time-dependent fields $\Psi_1^{\pm}$ when the assumed  boundary condition is I, so that the spectrum is given by (\ref{SpectrumSpinor1}), we find that
\begin{widetext}
\begin{align}
 \left. \Psi_1^{+} \right|_{r_\star\rightarrow -\infty}  &= \beta \,e^{-i\omega t} e^{-\gamma(c-1)r_\star}  \nonumber\\
   &= \beta \,e^{\mp i \sqrt{\mu^2 + q^2 Q_1^2 R_1^2 + L^2} (t-r_\star)}\,
e^{ (n+\frac{1}{2})\frac{t}{R_1}}   \, e^{  -( n + i  q Q_1 R_1^2 )\frac{r_\star}{R_1}  } \,,
\nonumber\\
 \left. \Psi_1^{+}\right|_{r_\star\rightarrow +\infty} &= \,
\beta  \frac{\Gamma(a+b-c)\Gamma(2-c)}{ \Gamma(a-c+1) \Gamma(b-c+1) } e^{-i\omega t} e^{\gamma(a+b-c)r_\star}
\nonumber\\
   &= \beta \frac{\Gamma(a+b-c)\Gamma(2-c)}{ \Gamma(a-c+1) \Gamma(b-c+1) } \,e^{\mp i \sqrt{\mu^2 + q^2 Q_1^2 R_1^2 + L^2} (t+r_\star)}\,
e^{ (n+\frac{1}{2})\frac{t}{R_1}}   \, e^{  ( n+1 - i  q Q_1 R_1^2 )\frac{r_\star}{R_1}  }  \,,
\nonumber\\
  \label{Psi-Asymptotic} \\
\left. \Psi_1^{-} \right|_{r_\star\rightarrow -\infty}  &= \alpha \,e^{-i\omega t} e^{\gamma(c-1)r_\star}
\nonumber\\
   &= \alpha \,e^{\mp i \sqrt{\mu^2 + q^2 Q_1^2 R_1^2 + L^2} (t-r_\star)}\,
e^{ (n+\frac{1}{2})\frac{t}{R_1}}   \, e^{  -( n+ 1 + i  q Q_1 R_1^2 )\frac{r_\star}{R_1}  } \,,
\nonumber\\
 \left. \Psi_1^{-}\right|_{r_\star\rightarrow +\infty} &= \,
  \alpha\, \frac{\Gamma(a+b-c)\Gamma(c)}{ \Gamma(a) \Gamma(b) }  e^{-i\omega t} e^{\gamma(a+b-c)r_\star}
\nonumber\\
   &= \alpha \, \frac{\Gamma(a+b-c)\Gamma(c)}{ \Gamma(a) \Gamma(b) } \,e^{\mp i \sqrt{\mu^2 + q^2 Q_1^2 R_1^2 + L^2} (t+r_\star)}\,
e^{ (n+\frac{1}{2})\frac{t}{R_1}}   \, e^{  ( n  - i  q Q_1 R_1^2 )\frac{r_\star}{R_1}  }  \,. \nonumber
\end{align}
\end{widetext}
Looking at these asymptotic forms, two features stand out: (i) the solutions do not represent progressive waves moving to the right or left, as we should demand from the boundary condition; (ii) since $n$ is real and positive, both fields $\Psi^{\pm}$ diverge  exponentially at the boundaries. It seems that something is wrong. Nevertheless, this impression comes from the fact that we looking at the fields themselves instead of analysing the conserved current that describes the flux of Dirac particles.

%
%
%

The conserved current associated to the Dirac field interacting with the background electromagnetic field is $J^\alpha = \bar{\Psi}\gamma^\alpha \Psi$, where $\bar{\Psi}$ stands for  the adjoint of $\Psi$, which for the representation adopted at Ref. \cite{Joas} is given by
\begin{equation*}
  \bar{\Psi} =  \Psi^{\dagger}\, (\sigma_3\otimes\sigma_3\otimes \cdots \otimes\sigma_3 ) \,,
\end{equation*}
see Ref. \cite{DissertaçãoJoas}. In particular, the current along the radial direction is given by:
\begin{align*}
  J^{\tilde{1}} &= (e_{\tilde{1}})_\alpha \, J^\alpha \\
&= \Psi_1(t,r_\star)^{\dagger}\sigma_3\sigma_2 \Psi_1(t,r_\star) \times (\textrm{Angular Part}).
\end{align*}
Thus, ignoring the multiplicative factor coming from angular dependence, it follows that the radial current is given by
\begin{equation}\label{J1-v2}
  J^{\tilde{1}} = \textrm{Re}\{\, \Psi_1^{+} \, (\Psi_1^{-})^{*} \, \} \,,
\end{equation}
where $*$ stands for complex conjugation and Re$\{\cdots\}$ takes the real part of its argument. Then, inserting the asymptotic forms (\ref{Psi-Asymptotic}) into Eq. (\ref{J1-v2}), lead us to the following asymptotic behaviour for the current when the boundary condition is I:
\begin{equation}\label{J1Asymptotic}
      \begin{array}{ll}
     \left. J^{\tilde{1}} \right|_{r_\star\rightarrow -\infty}  & \sim\, e^{(2n+1)\frac{(t-r_\star)}{R_1}} \,, \\
\left. J^{\tilde{1}} \right|_{r_\star\rightarrow +\infty}    &\sim \,  e^{(2n+1)\frac{(t+r_\star)}{R_1}} \,.
    \end{array}
\end{equation}
Thus, since the dependence of $J^{\tilde{1}}$ on the coordinates $t$ and $r_\star$ occur just through combinations $(t-r_\star)$ and $(t+r_\star)$, it follows that $J^{\tilde{1}}$ becomes a progressive wave at boundaries. In particular, at $r_\star\rightarrow -\infty$ the flux of particles is in the direction of increasing $r_\star$, while at $r_\star\rightarrow +\infty$ the flux of particles is in the direction of decreasing $r_\star$, which is in perfect accordance with the boundary condition I.

\begin{figure}
  \centering
  \includegraphics[width=5cm]{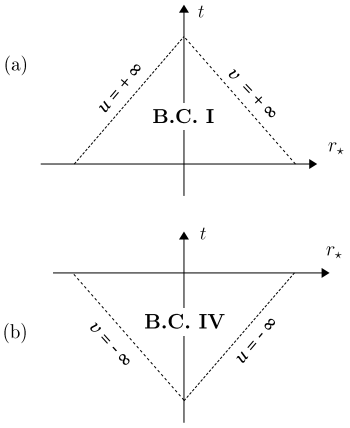}
  \caption{The dashed lines denote the region where the current is ill-defined. Part (a) corresponds to boundary condition I (B.C. I), while part (b) is corresponds to boundary condition IV (B.C. IV).  }\label{FigBC}
\end{figure}

From the asymptotic behaviour shown at Eq. (\ref{J1Asymptotic}), one could conclude that current $J^{\tilde{1}}$ diverges exponentially at the boundaries. However, this can be circumvented for arbitrarily large negative times. Indeed, defining the null coordinates
\begin{equation*}
  u = t-r_\star \;\textrm{ and } \; v = t+r_\star \,,
\end{equation*}
we see, from Eq. (\ref{J1Asymptotic}), that for $r_\star\rightarrow -\infty$ the current $J^{\tilde{1}}$  is ill-defined at $u\rightarrow + \infty$, but well-defined elsewhere. On the other hand, for $r_\star\rightarrow +\infty$ the current is diverges at $v\rightarrow + \infty$, while it is well-defined in other regions of the spacetime. This means that, for the boundary condition I, the current is ill-defined at the future null infinity, but it is non-divergent elsewhere, as depicted at the part (a) of Fig. \ref{FigBC}.

Analogously, computing the asymptotic form of the current $J^{\tilde{1}}$ for the solution corresponding to the boundary condition IV, we find that
\begin{equation}\label{J1AsymptoticIV}
      \begin{array}{ll}
     \left. J^{\tilde{1}} \right|_{r_\star\rightarrow -\infty}  & \sim\, e^{-(2n+1)\frac{(t+r_\star)}{R_1}} = e^{-(2n+1)\frac{v}{R_1}}  \,, \\
\left. J^{\tilde{1}} \right|_{r_\star\rightarrow +\infty}    &\sim \,  e^{-(2n+1)\frac{(t-r_\star)}{R_1}} = e^{-(2n+1)\frac{u}{R_1}} \,.
    \end{array}
\end{equation}
Thus, for the boundary condition IV, the current is divergent for $u\rightarrow - \infty$ and $v\rightarrow - \infty$. In other words, the current is ill-defined at the past null infinity, but it is well-defined elsewhere. The part (b) of Fig. \ref{FigBC} shows the region where the current is divergent for the boundary condition IV.

The very same behaviour is found for the current of the scalar field, whose conserved current is defined by
\begin{equation*}
  \mathcal{J}_\mu = \textrm{Im}\{\, \Phi \,\partial_\mu  \Phi^*  \,\} \,.
\end{equation*}
Indeed, computing this current at the boundaries using the asymptotic form of the scalar field obeying the boundary condition I lead us to the following
current on the radial direction:
\begin{equation*}
      \begin{array}{ll}
     \left.  \mathcal{J}_{r_\star} \right|_{r_\star\rightarrow -\infty}  & \sim\, e^{(2n+1)\frac{(t-r_\star)}{R_1}} = e^{(2n+1)\frac{u}{R_1}}  \,, \\
\left. \mathcal{J}_{r_\star} \right|_{r_\star\rightarrow +\infty}    &\sim \,  e^{(2n+1)\frac{(t+r_\star)}{R_1}} = e^{(2n+1)\frac{v}{R_1}} \,.
    \end{array}
\end{equation*}
This is the same behaviour of the spinorial current for the boundary condition I, see Eq. (\ref{J1Asymptotic}). Likewise, when the adopted boundary condition is IV, the component $\mathcal{J}_{r_\star}$ has the same asymptotic form of current $J^{\tilde{1}}$ at Eq. (\ref{J1AsymptoticIV}).

The fact that there exists regions of the spacetime where the physical current is ill-defined should not come as a surprise.
Indeed, the reason why the solutions for the boundary conditions I and IV are called quasinormal modes instead of normal modes is that the spectrum of allowed frequencies has also an imaginary part. Therefore, the time dependence of the fields, $e^{-i\omega t}$, blows up at $t\rightarrow \infty$ when $\textrm{Im}\{\omega\} > 0$, whereas it diverges at $t\rightarrow -\infty$ for $\textrm{Im}\{\omega\} < 0$. Generally, the QNMs are thought as states that are not existent at all times, rather they are excitations that occur at a particular time interval. In particular, they do not form a complete basis for the space of solutions of considered field equation \cite{Nollert99}.

\section{Conclusions and Perspectives}\label{Sec.Conclusion}

In this article we have investigated the perturbations on a scalar field, an spinorial field and a spin 1 gauge field propagating in a generalized version of the charged Nariai spacetime. Besides the separability of the degrees of freedom of these perturbations, which has been attained at Secs. \ref{Sec.Scalar}, \ref{Sec.Maxwell}, and \ref{Sec.Spinor}, one interesting feature of this background is that the perturbations can also be analytically integrated. They all obey a Schr\"{o}dinger-like equation with an integrable potential, with the solution given in terms of hypergeometric functions, as shown in Sec. \ref{Sec.Potential}. This is a valuable property, since even the perturbation potential associated to the humble Schwarzschild background is non-integrable, in spite of being separable.

Here, we have also investigated the quasinormal modes (QNMs) of these test fields.  We have seen that, from the causal point of view, the natural boundary conditions to be imposed to the perturbations are II and III of Fig. \ref{FigBoundCond}, but they do not lead to QNMs. On the other hand,  the \textit{ad hoc} boundary  conditions I and IV of Fig. \ref{FigBoundCond} do allow QNMs. For the convenience of the reader, the obtained quasinormal frequencies (QNFs) are summarized at table \ref{TableQNM} (below). Analysing this table, it is interesting noting that the imaginary parts of the QNFs, which represent the decay rates, are generally the same for the three types of fields and they do not depend on any detail of the perturbation, rather they only hinge on the charges of the gravitational background, through the dependence on $R_1$. Differently, the real parts of the QNFs depend on the mass of the field and on the angular mode of the perturbations. Another fact worth pointing out is that while the fermionic field always has a real part on its QNF spectrum, meaning that it always oscillates, the bosonic fields can have purely imaginary QNM frequency. Indeed, due to the negative factor ``$-1/(4R_1^2)$'' inside the square root appearing at the bosonic spectrum, it follows that for small enough $R_1$, along with small enough mass and angular momentum, the argument of the square root can be negative, so that this term becomes imaginary.

Once we have integrated the perturbations of test fields in the higher-dimensional generalization of Nariai spacetime (\ref{nariai-metric}), as well as studied their boundary conditions, the next natural step is to consider the perturbations on the metric field. The research on the latter problem is still ongoing and, due to the great number of degrees of freedom in the gravitational field, shall be considered in a future work.

\begin{widetext}
\begin{center}
\begin{table}[h]
\begin{tabular}{c||c|c|c|c}
  \hline
  \textbf{Field} & $\bl{\omega_{\textrm{I}}}$  & $\;\bl{\omega_{\textrm{II}}}\;$ & \;$\bl{\omega_{\textrm{III}}}\;$ & $\bl{\omega_{\textrm{IV}}}$ \\ \hline  \hline
  Scalar &  $\pm \left[\mu^2  + \sum_{j}  \frac{\ell_{j}(\ell_j + 1)}{R_j^2}   - \frac{1}{4R_1^2} \right]^{1/2} + \frac{i}{2R_1} (2n+1)$
& $\times$ & $\times$ &
 $\pm \left[\mu^2  + \sum_{j}  \frac{\ell_{j}(\ell_j + 1)}{R_j^2}   - \frac{1}{4R_1^2} \right]^{1/2} - \frac{i}{2R_1} (2n+1)$
\\ \hline
  Spinorial & $\pm\, \left( \mu^2 + q^2 Q_1^2 R_1^2 + L^2 \right)^{1/2} + \frac{i}{2R_1}(2n+1)$ & $\times$  & $\times$ &
$\pm\, \left( \mu^2 + q^2 Q_1^2 R_1^2 + L^2 \right)^{1/2} - \frac{i}{2R_1}(2n+1)$
\\ \hline
  Maxwell & $\pm  \left[ \sum_{j}  \frac{\ell_{j}(\ell_j + 1)}{R_j^2}   - \frac{1}{4R_1^2} \right]^{1/2}  + \frac{i}{2R_1} (2n+1)$ & $\times$  &
$\times$ & $\pm  \left[ \sum_{j}  \frac{\ell_{j}(\ell_j + 1)}{R_j^2}   - \frac{1}{4R_1^2} \right]^{1/2}  - \frac{i}{2R_1} (2n+1)$  \\
  \hline
\end{tabular}
\caption{Allowed frequencies for the three types of matter fields considered here and for the four boundary conditions described in Fig. \ref{FigBoundCond}. Above, $\omega_{\textrm{I}}$ stands for the frequencies when the boundary condition is I, for instance, while $\times$ indicates the absence of QNFs. }\label{TableQNM}
\end{table}
\end{center}
\end{widetext}

\begin{acknowledgments}
C. B. would like to thank Conselho Nacional de Desenvolvimento Cient\'{\i}fico e Tecnol\'ogico (CNPq) for the partial financial support through the research productivity fellowship. Likewise,  C. B. thanks Universidade Federal de Pernambuco for the funding through Qualis A project.  J. V. thanks CNPq for the financial support.
\end{acknowledgments}

\appendix

\section{Ansatz for the Separation of Maxwell's Equation}\label{Appendix.Maxwell}

Just as in a spherically symmetrical problem it is useful to expand the angular dependence of a scalar field in terms of spherical harmonics, in the case of a vector field it is useful to make the expansion in terms the so-called vector spherical harmonics. The latter objects are given by
\begin{equation*}
  \left\{
     \begin{array}{ll}
       \vec{\mathcal{E}}_{1,\ell m} = \frac{1}{r}\vec{r}\,\, Y_{\ell}^{m}(\theta,\phi) \,, \\
       \vec{\mathcal{E}}_{2,\ell m} = \vec{r} \times \vec{\nabla} Y_{\ell}^{m}(\theta,\phi) \,, \\
        \vec{\mathcal{E}}_{3,\ell m} = r\, \vec{\nabla} Y_{\ell}^{m} (\theta,\phi)   \,,
     \end{array}
   \right.
\end{equation*}
where $\{r,\theta,\phi\}$ is a spherical coordinate system in $\mathbb{R}^3$. Since these three vector fields are orthogonal to each other, it follows that they are linearly independent and, therefore, form a frame for the space of vector fields in  $\mathbb{R}^3$. Thus, in a spherically symmetric problem it is natural to expand vector fields $\vec{A}$ in terms of this base,
\begin{equation}\label{A}
  \vec{A} = A^1(r) \,\vec{\mathcal{E}}_{1,\ell m} + A^2(r) \,\vec{\mathcal{E}}_{2,\ell m} + A^3(r) \,\vec{\mathcal{E}}_{3,\ell m}  \,.
\end{equation}
Using the expression for the gradient in spherical coordinates, it follows that the vector spherical harmonics are given by
\begin{align}
 \vec{\mathcal{E}}_{1,\ell m} &=  Y_{\ell}^{m}(\theta,\phi)\,\hat{e}_r  \;,\;\; \nonumber \\
          \vec{\mathcal{E}}_{2,\ell m} &= \frac{-\,1}{\sin\theta}\partial_\phi Y_{\ell}^{m}(\theta,\phi)\, \,\hat{e}_\theta
          + \partial_\theta Y_{\ell}^{m}(\theta,\phi)\, \,\hat{e}_\phi \,, \;\;\;  \nonumber \\
           \vec{\mathcal{E}}_{3,\ell m} &=  \partial_\theta Y_{\ell}^{m}(\theta,\phi)\, \,\hat{e}_\theta
          + \frac{1}{\sin\theta}\partial_\phi Y_{\ell}^{m}(\theta,\phi)\, \,\hat{e}_\phi     \nonumber
\end{align}
where $\{\hat{e}_r,\hat{e}_\theta,\hat{e}_\phi\}$ is the orthonormal frame associated to the spherical coordinates $\{r,\theta,\phi\}$. More precisely, their connection with coordinate frame is the following:
\begin{equation*}
  \hat{e}_r = \partial_r \;\;,\;\;  \hat{e}_\theta = \frac{1}{r}\partial_\theta \;\;,\;\; \hat{e}_\phi = \frac{1}{r \sin\theta}\partial_\phi \,.
\end{equation*}
Thus, the generic vector field $\vec{A}$ of Eq. (\ref{A}) is written as
\begin{align*}
  \vec{A} =& A^1\,Y_{\ell}^{m}\,\hat{e}_r +  \left( A^3 \partial_\theta Y_{\ell}^{m} - A^2\frac{\partial_\phi Y_{\ell}^{m}}{\sin\theta} \right)\hat{e}_\theta \\
&+ \left( A^3  \frac{\partial_\phi Y_{\ell}^{m}}{\sin\theta} + A^2\partial_\theta Y_{\ell}^{m} \right) \hat{e}_\phi\,.
\end{align*}

Likewise, in a spherically symmetric problem, a 1-form $\bl{\mathcal{A}}$ is conveniently expanded in the following way
\begin{align*}
  \bl{\mathcal{A}} = & A^1\,Y_{\ell}^{m}\,\hat{e}^r +
\left( A^3 \partial_\theta Y_{\ell}^{m} - A^2\frac{\partial_\phi Y_{\ell}^{m} }{\sin\theta} \right)\hat{e}^\theta \\
& + \left( A^3  \frac{\partial_\phi Y_{\ell}^{m} }{\sin\theta} + A^2\partial_\theta Y_{\ell}^{m}  \right) \hat{e}^\phi\,,
\end{align*}
where $\{\hat{e}^r,\hat{e}^\theta,\hat{e}^\phi\}$ stands for the frame of 1-forms that is dual to the frame of vector fields $\{\hat{e}_r,\hat{e}_\theta,\hat{e}_\phi\}$, namely $\hat{e}^{a}(\hat{e}_{b})=\delta^a_b$. This frame is related to the coordinate frame $\{dr,d\theta,d\phi\}$ as follows:
\begin{equation*}
  \hat{e}^r = dr\,,\; \hat{e}^\theta=r d\theta \,,\; \hat{e}^\phi=r \sin\theta d\phi \,,
\end{equation*}
so that the line element of $\mathbb{R}^3$ is written, in spherical coordinates, as $ds^2 =  (\hat{e}^r)^2 + (\hat{e}^\theta)^2 +(\hat{e}^\phi)^2$. Thus,  the expansion of the 1-form can be written as:
\begin{align}
   \bl{\mathcal{A}} = &  \tilde{A}^1 \,Y_{\ell}^{m}\,dr  - \tilde{A}^2  \, \frac{\partial_\phi Y_{\ell}^{m}}{\sin\theta} \,d\theta \nonumber\\
& + \tilde{A}^2   \, \sin\theta \partial_\theta Y_{\ell}^{m} \, d\phi + d\left( rA^3  Y_{\ell}^{m}  \right) \,, \label{A-gauge}
\end{align}
 where $\tilde{A}^1 = A^1 - A^3 - r \frac{dA^3}{dr}$ and $\tilde{A}^2=r A^2$ are functions of the radial coordinate $r$. In a $U(1)$ gauge field theory, we can ignore the last term in the previous equation, since an exact differential can be eliminated by a gauge transformation. Thus, we can say that a natural ansatz for a 1-form gauge field in a problem with spherical symmetry is:
\begin{align*}
   \bl{\mathcal{A}}_{\ell,m} = &  \tilde{A}^1(r) \,Y_{\ell}^{m}(\theta,\phi)\,dr
- \tilde{A}^2(r)  \, \frac{\partial_\phi Y_{\ell}^{m}(\theta,\phi)}{\sin\theta} \,d\theta \\
& + \tilde{A}^2(r)   \, \sin\theta\, \partial_\theta Y_{\ell}^{m}(\theta,\phi) \, d\phi   \,.
\end{align*}

In the problem considered in the present article, the spacetime is the direct product of $dS_2$ with several spheres, so that we have spherical symmetry in each of these spheres. Thus, the ansatz for the gauge field which is in agreement with such symmetries is
\begin{align}
& \bl{\mathcal{A}} =   e^{-i\omega t} \bigg[ \,\tilde{A}^0\,\mathcal{Y}_{\ell,m} \,dt + \tilde{A}^1\,\mathcal{Y}_{\ell,m} \,dr_\star   \label{AnsatzA}\\
 &  - \sum_{j=2}^{d} \tilde{A}^j   \, \left( \frac{\partial_{\phi_j} \mathcal{Y}_{\ell,m} }{\sin\theta_j} \,d\theta_j
  -     \sin\theta_j \,\partial_{\theta_j} \mathcal{Y}_{\ell,m}\, d\phi_j \right) \bigg]\,, \nonumber
\end{align}
with $\tilde{A}^0 = \tilde{A}^0(r_\star)$, $\tilde{A}^1= \tilde{A}^1(r_\star)$ and $\tilde{A}^j= \tilde{A}^j(r_\star)$ being arbitrary functions of the tortoise coordinate $r_\star$, while $\mathcal{Y}_{LM}$ is the function of the angular coordinates defined in Eq. \eqref{YLM}. The collective subindex $\ell,m$ is an abbreviation for the set of indices $\{\ell_2, m_2, \ell_3, m_3,\cdots, \ell_d, m_d\}$.

Actually, it is important to point out that although we have been able to eliminate the last term in Eq. \eqref{A-gauge} by means of a gauge transformation, the annihilation of this degree of freedom cannot be done in the higher-dimensional case considered here. Indeed, when the symmetry of the space is a product of spherical symmetries, the eliminated term represents a relevant degree of freedom to form the most general ansatz, namely the components $A^3(r)$ (one for each sphere) cannot be eliminated by a gauge transformation. Nevertheless, working out the solutions of Maxwell's equation, taking into account these extra degrees of freedom, we find that these functions also satisfy Eq. \eqref{WaveEqMaxwell}, so that the spectrum of these extra degrees of freedom  is the same of the other components of the Maxwell field. For this reason, although not the most general, it is enough to consider the simpler ansatz \eqref{AnsatzA}.


\section{The Angular Part of the Dirac Equation}\label{Appendix.Angular}

The separation of the Dirac equation for spacetimes that are the direct product of 2-dimensional spaces has been done at Ref. \cite{Joas}. Applying Eq. ($27$) of the latter reference to the spacetime considered here, lead us to the conclusion that the angular parts of the spinorial field obey the following differential equation
\begin{align*}
 & \left[ i s_{j} \left(\frac{1}{\text{sin} \theta_{j}} \partial_{\phi_{j}} - iq Q_{j}R_{j}^{2} \text{cot} \theta_{j}\right) \right.  \nonumber \\
& \left. + \left(\partial_{\theta_{j}} + \frac{1}{2} \text{cot} \theta_{j} \right)  \right]\Psi_{j}^{-s_{j}}
= R_{j}(c_{j} - s_{j}\,c_{j-1} )\,\Psi_{j}^{s_{j}} \,,
\end{align*}
where $c_j$ are separation constants. Now, inasmuch as the coefficients in the above equation are independents of the coordinate $\phi_{j}$, which stems from the fact that the vector field $\partial_{\phi_{j}}$ is a Killing vector of our metric, we can assume the following angular decomposition for the fields
$$
\Psi^{s_{j}}_{j}(\phi_{j}, \theta_{j}) \,=\, e^{-i\omega_{j}\phi_{j}}\,\psi^{s_{j}}_{j}(\theta_{j}) \,.
$$
Then, inserting this decomposition into the above differential equation, we are left with the following:
\begin{align}
 &\left[\frac{d}{d\theta_{j}} + \frac{1}{2}\,\text{cot}\,\theta_{j} + s_{j} \left(\frac{\omega_{j}}{\text{sin}\,\theta_{j}} + q Q_{j}R_{j}^{2}\,\text{cot}\,\theta_{j}\right) \right]\psi_{j}^{-s_{j}} \nonumber\\
 &= R_{j}(c_{j} - s_{j}\,c_{j-1} )\,\psi_{j}^{s_{j}} \,. \label{PAED2}
\end{align}

Now, let us write this Eq. (\ref{PAED2}) in a more convenient form. In order to accomplish this,  instead of using $\{c_j\}$, let us use the parameters $\{\lambda_j\}$ defined by
$$
\frac{\lambda_{j}}{R_{j}} \equiv \sqrt{c_{j-1}^{2} - c_{j}^{2}} \,.
$$
Inverting these expressions, we find that
$$
c_{j-1} \,=\,\sqrt{\frac{\lambda_{j}^{2}}{R_{j}^{2}} \,+\,\frac{\lambda_{j+1}^{2}}{R_{j+1}^{2}}\,+\, \ldots \,+\, \frac{\lambda_{d}^{2}}{R_{d}^{2}}} \,.
$$
In particular, the constant $L$ appearing at Eqs. (\ref{DifEqSpinor}) and (\ref{L}) should be identified with $c_1$. Next, if we introduce the parameter
$$
\zeta_{j} \,=\,\text{arctanh}(c_{j}/c_{j-1}) \,,
$$
it is a simple matter to prove that
$$
c_{j} \,=\,  \frac{\lambda_{j}}{R_{j}}\,\text{sinh}\,\zeta_{j} \quad , \quad c_{j-1} \,=\,  \frac{\lambda_{j}}{R_{j}}\,\text{cosh}\,\zeta_{j} \,.
$$
Moreover, using the above identities, one can also verify that $c_{j-1}$ and $c_{j}$ satisfy the relation
$$
c_{j} - s_{j}c_{j-1}\,=\,-s_{j}\lambda_{j}\,e^{-s_{j}\zeta_{j}}/R_{j} \,.
$$
In addition to this change of parameters, let us make the field redefinition
$$
\phi_{j}^{s_{j}} \,=\,e^{-s_{j}\zeta_{j}/2}\psi^{s_{j}}_{j} \,.
$$
In terms of $\phi_{j}^{s_{j}}$, Eq. \eqref{PAED2} reads:
\begin{align*}
 & \left[\frac{d}{d\theta_{j}} + \frac{1}{2}\,\text{cot}\,\theta_{j} - s_{j} \left(\frac{\omega_{j}}{\text{sin}\,\theta_{j}} + q Q_{j}R_{j}^{2}\,\text{cot}\,\theta_{j}\right) \right]\phi_{j}^{s_{j}} \\
&= s_{j}\lambda_{j}\,\phi_{j}^{-s_{j}} \,.
\end{align*}
In particular, when the magnetic charges of the background vanish, $Q_j=0$, we find
\begin{equation}
\left[\frac{d}{d\theta_{j}} + \frac{1}{2}\,\text{cot}\,\theta_{j} -  \frac{s_{j} \omega_{j}}{\text{sin}\,\theta_{j}} \right]\phi_{j}^{s_{j}} = s_{j} \lambda_{j}\,\phi_{j}^{-s_{j}} \,,
\end{equation}
which is the Dirac equation at the unit 2-sphere. The latter equation admits regular analytical solutions only when the eigenvalues $\lambda_{j}$ are  nonzero integers \cite{Camporesi:1995fb,Abrikosov:2001nj}, $\lambda_{j} = \pm 1, \pm 2, \ldots $.



\end{document}